\documentclass{birkjour}
 \theoremstyle{definition}
 
 \theoremstyle{remark}

 \numberwithin{equation}{section}

\begin{document}
\title{Quantum theory of time perception: phases, clocks and quantum algebra}

\author{Rukhsan Ul Haq}

\address{%
Theoretical Sciences Unit\\
Jawaharlal Nehru Center for Advanced Scientific Research\\
Jakkur Bangalore India}

\email{mrhaq@jncasr.ac.in}

\author{Shalini Harkar}

\address{
Center for Nanoscience\\
Indian Institute of Science\\
Bangalore India}

\email{hshalini@cense.iisc.ernet.in}


\date{}

\begin{abstract}
Experience of time is one of the primordial human experiences which is
deeply tied to human consciousness. But depsite this intimate relation of
time with human conscious experience, time has proved to be very elusive.
Particularly in physics, though there is already some understanding of
time, there are still so many paradoxes that plague this understanding. In
this paper we take rather a different route to question of time. 
We first attempt to come up with a theoretical understanding of time perception. 
Quite interestingly we find that quantum theory provides
an algebraic formulation within which we can understand some 
essential aspects of time perception by human mind. We then ask whether
a similar formalism can furnish the understanding of time as well and 
find connections of our formulation of time with similar works by other researchers.
Our underlying approach to question of time has been inspired by R. W. Hamilton
who considers algebra as science of pure time. Hence our work has an extensive
algebraic flavor. Our work also incorporates another approach based on Kauffman's
iterant algebra which relates time to underlying recursions and oscillations. 
We believe that our work will initiate more investigations in this direction.

\end{abstract}

\maketitle

\section{Introduction}
Space and time are integral part of our conscious experience.Our cognition of 
 outside world happens through the mental construction of spatial and temporal
 structures. They are the two basic varieties of our cognitive apparatus. It is no 
 wonder then that space and time form the most fundamental concepts of any theory
 of the universe. Space and time have always been the focus of philosophers,mathematicians and scientists in general, right from the very
 beginning of human civilizations. Because of the fact that space and time are
 intimately related to human psyche they not only arise in philosophical and
 scientific discourses rather they understandably also form the fabric for 
 the artistic and religious creativity. In mystic circles, space and time are 
 considered very sacred because they also form the underlying arena over which
 unfolds the celestial drama of divine creativity. It is within space and time
 that human intellect catches the glimpses of divine creativity.\\
    For Greeks space was very sacred because it represented an eternal principle  which witnessed all the change and flux of the world and itself remaining aloof from it. So
 space gave them sense of permanence among the transitory world of change and flux.
 Space attained religious significance. Historically Greeks were the first people who did a very systematic and
 logically very coherent study of space and hence developed the geometry now
 known as Euclidean geometry. Euclid compiled the geometry in thirteen volumes
 and it the same geometry which is taught to all of us in schools. It should be 
 noted that it were Greeks once again who developed the first logical system 
 called Aristotelian logic and it is exactly the same logic which is used even now
 by all of us to settle our daily disputes and it is the same logic on which our computers also work. So we use Aristotelian logic without 
 actually realizing it. The Greek geometry was modelled on their logic so Euclid wrote geometry
 as a logical system using an axiomatic method of presentation. Roger Penrose,
 a very renowned contemporary theoretical physicist, reckons Euclidean geometry
 as one of the first rate theories of the world. Greeks also developed projective
 geometry which found very enormous applications in renaissance art and painting.
 Projective geometry has emerged again as the  geometry of the state space in quantum theory. Though
 Greeks did  quite an impressive work in studying space,they could not make any
 advance in understanding time. In fact the dominant world view of Greeks was 
 static with geometry lying at its foundation. The study of time is called dynamics
 and the principles which include time as an important factor are called dynamic
 principles. So Greeks only could study static and changeless features of the world
 and first dynamical theory had to wait for the genius of Isaac Newton till 17th
 century.
 But with Newton once
 again the maximum he could do or say about time was to consider it to belong to
 divine category and hence was taken to be absolute,above all change. It was only
 at the turn of last century that first ever theory was put forward by Albert 
 Einstein who treated space and time on an equal basis with both of them as 
 relative individually but taken together they form the fundamental element of
 geometry which he later on showed to be intimately related to dynamics.
 It was for the first time in  history of science that geometry and dynamics along with matter
 came so close that they made one fabric out of which arise the spectacular 
 phenomena
 of the world and ever since physicists and mathematicians have been struggling 
 to come up with a mathematical structure which will include all fundamental
 forces of the world. But the problem which Einstein left us in, is that his 
 spacetime is not Euclidean but is hyberbolic. That means in his geometry of the
 world though space is real but time is imaginary. In this synthesis though space
and time have come closer than ever but are distanced from our conscious 
experience. In our conscious experience we see space as three dimensional Euclidean
  geometry and time flowing as if on an independent axis. So with all its glorious
  and historic triumphs, physics still could not resolve paradox of space and time.
  Time has proved to be very elusive to mathematical analysis and yet all
  of us in our waking consciousness  experience  time and it forms basic thread
  of our consciousness. The time which physicists talk about in their theories is 
  symmetric 
  between past and future and yet all of us know very painfully that our past can not be
  exchanged with our future. So one of the greatest problems which theoretical
  physics is facing is actually a very fundamental question: "What is time?"
In this paper  we take a practical approach to this question. We first develop a formulation to understand time perception and then  consider the more general question. What is very interesting is that  quantum theory provides  a rich formulation within which we address these questions. Along the way we find an interesting algebra which we believe can provide the algebraic setting for the understanding of physical,biological and psychological aspects of time.
 

We first of all note that psychological time is different from physical time, where by latter we mean the time which is used in dynamical laws of physics like the one in Newton's equations. In quantum theory there is one difference though, that time gets multiplied by imaginary unit so that dynamics is unitary. That is how Schrodinger equation is different from diffusion equation even though both are first order in time.
The dynamical laws of physics have manifest time reversal symmetry while as psychological time has categorical distinction between past and future which can not be replaced. Another problem with physical time is that it is real variable and hence moves on time axis in a serial manner while as the way time is measured and also experienced in biological systems is periodic and hence the arithmetic of time becomes modular. Modular nature of arithmetic of time is one essential feature of time which is missed out in physics but plays important role in time perception. This fact opens up the relation between question of time and number theory\cite{Schroder}. It is interesting to note that in quantum field theory and quantum statistical mechanics when time is mapped to temperature it becomes periodic.\\
 These are essential differences of psycological time with physical time and any theory of time has to explain them. Now from the research in cognitive psychology we also know that psychological time has two more interesting features. One is that time perception takes place at logarithmic scale \cite{Planat1}and $\frac{1}{f}$ noise is the ubiquitous in human cognition and hence also in time perception\cite{Planat2}\cite{Simon}. In this paper we take an algebraic approach to time because with Hamilton\cite{Hamilton} we take algebra as science of time. However there have been very illuminating geometric approaches to understand the nature of time. However we will not dwell on them here and just mention that the algebraic structures we use have their geometrical side as well, particularly they are deeply related to projective geometry which is what has been used in to understand the geometry of time\cite{Saniga}.

\section{$Z_{n}$ clocks and quantum algebra}
We begin with  a clock which has only two time values: two values of square root of unity. We call this clock a $Z_{2}$ clock. $Z_{2}$ here refers to to the set ${1,-1}$ which are values not only of square root of unity but can also refer to parity of Majorana fermion or to reflection of a spinor. In this section we will always refer to two values of square root of unity.
We introduce two operators or matrices. 
\begin{equation}
e=	\left(\begin{array}{cc}     
			1&0\\
			0&-1
			\end{array}\right)
\end{equation}
\begin{equation}
\eta=	\left(\begin{array}{cc}     
			0&1\\
			1&0
			\end{array}\right)
\end{equation}
We call $e$ as clock matrix and $\eta$ as shift matrix. Clock matrix has values of time as its eigenvalues while as shift matrix changes the time states into one another and hence shifts the time on the $Z_{2}$ clock. 
What is very interesting is that $e$ and $\eta$ obey Clifford algebra which though integral to the physics,geometry and topology has rarely been seen to have any connection to the question of time. So we would like to understand this connection better. To do that we take an approach based on what has been called as iterants. The imaginary unit is a kind of clock because it keeps on oscillating between two values which are square roots of unity. So if we take a time series of this clock ,what we will get at any instant is something of this sort:
$...+1-1+1-1+1-1+1-1+1-1....$ which after one time shift will become:$...-1+1-1+1-1+1-1+1...$. Each one of these are called iterants and there is an algebra which has been called iterant algebra. For the case of only two values iterant algebra is same as the algebra of $Z_{2}$ clock. In fact in iterant we have the serial version of what is actually happening in clock that there are these discrete time values and there is a shift operator which is permuting  these values. Due to this fact we see that iterants are  cyclical and modular. However iterant approach is very interesting because  it  relates time to the recursive processes. The oscillations present in iterants are due to the logical paradox. We borrow from Kauffman who has developed the iterant algebra, a very succint  description of iterants\cite{Kauffman1}:"The simplest discrete system  corresponds directly to the  square root of  minus one,  when the
square root of minus one is seen as an oscillation between plus and minus one. This way thinking about the square root of minus one as an iterant
is explained below. More generally, by starting
with a discrete time series of positions, one has immediately a non-commutativity of observations since the measurement of velocity involves the tick of the clock and the measurment of position does not demand the tick of the clock. Commutators that arise
from discrete observation generate a non-commutative calculus,  and this calculus leads to a  generalization of standard advanced
calculus in terms of a non-commutative world.  In a non-commutative world, all derivatives are
represented by commutators.
In this view, distinction and process arising from distinction is at the base of the world. Distinctions are elemental bits of awareness. The world is composed not of things but processes and
observations. We will discuss how basic Clifford algebra comes from very elementary processes
like an alternation of
$+
−
+
−
+
−···$
and the fact that one can think of
$√-1$
itself as a temporal
iterant, a product of an
$ǫ$
and an
$\eta$
where the
$ǫ$
is the
$+
−
+
−
+
−···$
and the
$\eta$
is a time shift
operator.  Clifford algebra is at the base of the world!" 
 What needs to be noted over here is that our clock algebra has all the essential properties of time incorporated right from the beginning: time is cyclical as it runs on a clock; arithmetic of time is modular and the algebra of time is Clifford algebra which is line with Hamilton's quaternion theory of time. Since our clock algebra is essentially iterant algebra so it is all the temporal aspects of iterants embedded.

Our $Z_{2}$ clock algebra is also related to Pauli matrices and that is surprising in the sense that Pauli matrices occur in the physics of spin. Though they also satisfy Clifford algebra however they do not seem to be related to time\cite{Rukhsan}. But we should note that Pauli matrices are related to quaternions which were introduced as the algebra of time.
In the canonical basis,$e=\sigma_{z}$,$\eta=\sigma_{x}$ and $\sigma_{y}=ie\eta$.


\subsection{$Z_{3}$ Clock}
In this section we take cube roots of unity,{1,$\omega$,$\omega^{2}$} as the time values of our Clock which we now call as $Z_{3}$ Clock. Once again we have the shift operator which permutes  the time values.
Then clock and shift matrices of $Z_{3}$ clock are:
\begin{equation}
e=	\left(\begin{array}{ccc}     
			1&0&0\\
			0&\omega&0\\
			0&0&\omega^{2}
			\end{array}\right)
\end{equation}
\begin{equation}
\eta=	\left(\begin{array}{ccc}     
			0&1&0\\
			0&0&1\\
			1&0&0
			\end{array}\right)
\end{equation}
But now $e$ and $\eta$ dont obey Clifford algebra. Rather they obey Generalized Clifford algebra.
\begin{equation}
e^{3}=\eta^{3}=1 \quad e\eta=\omega \eta e
\end{equation}
This algebra arises in many other physical situations like in Potts model and as quasiparticles in quantum Hall systems. Recently there has been more interest this parafermion algebra because of the importance of the parafermions in topological quantum computing. However what is very relevant to our work is that this algebra arises in discrete gauge theories as symmetry algebra.




\subsection{$Z_{n}$ Clock and time perception}
In earlier two sections we have already given the concrete examples of the Clock algebra. The generalized Clock algebra will have $nth$ roots of unity. Now the question which arises is :does our $Z_{n}$ clocks describe the time perception? To answer that question we have to recognise that $Z_{n}$ clock algebra is what has been used by Bost and Connes\cite{Planat3} to come up with a dynamical system which dependes logarithmically on number operator and hence on the phase operator as well. Planat has taken notice of this logarithmic dependence and using the number-phase uncertainty relation of quantum mechanics,he finds that Bost-Connes Hamiltonian gives quantum version of Fechner law and hence explains the time perception on logarithmic scale. Since the partition function of Bost-Connes system is given by Riemann zeta function so we get a random distribution of prime numbers. Interestingly it also explains the $\frac{1}{f}$ noise in time perception. Hence our $Z_{n}$ clock algebra gives us the way to understand time perception. 



\section{Quantum theory of time}
We have seen how quantum theory provides the formalism for understanding the time perception by human mind. Now we ask the broader question: does quantum theory provide us the theory of time as well?  We will begin with reviewing with what has been already understood by some of the pioneering people of quantum mechanics. David Deutsch finds that time is essentially quantum mechanical concept. Though he has his own reasons to believe so. We see that there are more mathematicallly sound reasons as well to understand how time is essentially quantum mechanical. Which can be understood from that fact that as Hamilton said that algebra is science of pure time and he had introduced quaternions for the same purpose. It is very interesting to note that it is only in quantum mechanics that Hamilton's view of algebra as science of pure time comes out fully. Quantum mechanics is algebraic though geometry is always there  but the radical departure of quantum mechanics was to renounce the phase space picture of classical mecahnics and base the whole theory on the algebraic theory of vector spaces and linear transformations. Quaternions  emerge again when we deal with time reversal symmetry in quantum theory. Quaternions are deeply related to time reversal symmetry. This fact can be taken as quantum mechanical manifestation of Hamilton's approach of algebra and quaternions in particular as the science of pure time. That makes time algebraic and quantum mechanical .Yet another very profound aspect of quantum nature of time is found in the way bosons and fermions "experience" time. For bosons/scalar particles time reversal is just a complex conjugation while as for fermions the transformation becomes quaternionic rotation. This aspect of time not only brings out the quantum nature of time out but it also related time to the quantum statistics which makes sense in view of spin-statistics theorm.



\section{Conclusions}
In this paper we began with pointing out how physical time is different than psychological time. In biological systems time is always periodic while as in physics time is taken as real number which flows on a real axis is serial manner. We emphasize that serial time is an artifact of mathematical formalism used in physics and does not correspond to the actual experience of time. There is a fundamental disparity between time variable in dynamical laws of physics and the actual measurement of time which is always associated with clocks and hence is periodic . Then there is another fundamental issue with the way time is treated in relativity and quantum theory. This problem is called "problem of time" in quantum gravity. Given the central role of the time in any dynamics we have made attempts to come up with a formalism which sheds new light upon the nature of time and particularly psychological aspects of it. We have taken rather a different approach. We first ask more practical question of how time is actually perceived by human mind and we have constructed Clock algebras to study the time perception. It is well known in psychology literature that time perception happens at logarithmic scale like the sound perception which takes place at decibel scale. Yet another interesting aspect of the time perception is the $\frac{1}{f}$ noise. We have proposed an approach which takes into the essential features of time and come up with a formalism which explains the psychological aspects of time perception. In this regard we make connection to Planat's very important work in this direction.  Then following the same themes we address the question of whether there is formalism which can incorporate the essential aspects of time into its very axiomatic base. These essential aspects include oscillation,periodicity,modularity and Clifford algebraic structure of time operator. Our approach is in line with many other approaches which have been able to shed some light on the nature of time. We particularly have discussed about Bohm-Hiley approach based on process algebra which is based on iterant algebra. Similarly the chronon approach of Finkelstein  brings out the Clifford algebraic aspects of time operator. The connection to the thermodynamic arrow of time can be made by Wick rotation which relates time to temperature. We mark this relation as one of the very profound aspects of time which brings dynamics and thermodynamics together. We bring out this relation between time(dynamics) and temperature(thermodymanics) by looking at Schrodinger equation and diffusion equation which are same in other differential aspects however lead to entirely different classes of dynamics. The difference lies in the way time enters into these equations. In quantun dynamics time has to get mutltiplied with imaginary unit while as diffusion equation absensce of imaginary unit leads to stochastic dynamics. Unitarity of quantum dynamics is tied to the imaginary unit is Schrodinger equation. We believe that this paper has shed new light at the profound aspects of time and has put forward an approach which can lead us the natural framework within we can bring together physical,biological and psychological aspects of time. 

\end{document}